\documentclass[preprint,prd,aps,showpacs,showkeys,nofootinbib]{revtex4}
\usepackage{graphicx}
\usepackage{dcolumn}
\usepackage{bm}
\usepackage{color}
\begin{document}

\title{Scalar neutrino dark matter in BLMSSM}
\author{Ming-Jie Zhang$^{1,2}$\footnote{1070102415@qq.com}, Shu-Min Zhao$^{1,2}$\footnote{zhaosm@hbu.edu.cn}, Xing-Xing Dong$^{1,2}$\footnote{dxx$\_$0304@163.com}, Zhong-Jun Yang$^{3}$, Tai-Fu Feng$^{1,2,3}$}

\affiliation{$^1$ Department of Physics, Hebei University, Baoding 071002, China}
\affiliation{$^2$ Key Laboratory of High-precision Computation and Application of Quantum Field Theory of Hebei Province, Baoding 071002, China}
\affiliation{$^3$ Department of Physics, Chongqing University, Chongqing 401331, China}
\date{\today}

\begin{abstract}
  BLMSSM is the extension of the minimal supersymmetric standard model(MSSM). Its local gauge group is $SU(3)_C \times SU(2)_L \times U(1)_Y \times U(1)_B \times U(1)_L$. Supposing the lightest scalar neutrino is dark matter candidate, we study the relic density and the spin independent cross section of sneutrino scattering off nucleon. We calculate the numerical results in detail and find suitable parameter space. The numerical discussion can confine the parameter space and provide a reference for dark matter research.
\end{abstract}

\keywords{dark matter, sneutrino, supersymmetry}

\maketitle

\section{introduction}
Since it was first proposed in the 1920s, the existence of dark matter has been confirmed by many observations and experiments. In particular, the explanations of galaxy rotation curves\cite{dm1,dm2}, gravitational lensing and cosmic microwave background radiation have made dark matter widely
accepted by many physicists. People turn more attention to dark matter research.

 The neutrinos in SM  have hot dark matter characters, but
 the large scale structure of the universe supports that cold dark matter is dominant.
 So, SM can not provide cold dark matter candidate. Though SM has achieved great success with the detection of 125 GeV Higgs boson, it has some shortcomings,
 such as the hierarchy problem, CP-violating problem and neutrino with zero mass.
 Therefore, we consider to extend the SM, and the MSSM\cite{mssm1,mssm2,mssm3} is one choice of many supersymmetric models. However, the MSSM still fails to explain neutrino oscillation experiments
 and its reasonable parameter space constrained by the experiments turns smaller and smaller. So, we need to  expand the MSSM.

 The BLMSSM\cite{BL1,BL2,BL3,BL4}  is the extension of the MSSM with local gauged B
and L, whose local gauge group is $SU(3)_C\times SU(2)_L\times U(1)_Y \times U(1)_B\times U(1)_L$
and  spontaneously broken at TeV scale. The local gauged B
can explain the asymmetry of matter-antimatter in the universe. At the same time, the local gauged L can improve the lepton
flavor violating effect, and give light neutrino tiny mass with right handed neutrino through see-saw mechanism.
The little hierarchy problem in MSSM is relieved in BLMSSSM by the exotic quarks and exotic leptons
which are introduced to eliminate the gauge anomalies. BLMSSM can provide new dark matter candidate beyond MSSM.
 So, we choose to use the BLMSSM to study dark matter.

In the current dark matter research, there are many dark matter candidates. They include massive compact
halo objects(MACHOs)\cite{machos1,machos2,machos3}, primordial black holes, neutrinos, axion and weakly-interacting massive particles(WIMPs)\cite{wimps1,wimps2,wimps3,wimps4,wimps5}. They all satisfy or partially satisfy: no electric charge and no color charge, remaining stable and possessing a log life\cite{dm3,dm4}.
 These characters could explain the observed large scale structure of the universe. Then cold dark matter is favorite.

In MSSM, the scalar neutrinos are just left-handed and  interact with gauge bosons at tree level.
It can not satisfy the constraints from the relic density or direct detection experiments because of large cross section. The introduced right-handed neutrinos
are inactive and stable. The sneutrinos are electric and color neutral. If the lightest mass eigenstate of sneutrino mass squared matrix is the
lightest supersymmetric particle (LSP), and its dominant element is the right-handed sneutrino, it will be a good dark matter candidate. The reason is that
it can easily satisfy the  experimental restrictions from dark matter direct detection and give reasonable
relic density. Eventually, the lightest scalar neutrino is adopted as the candidate of dark matter in this paper. There are other studies of sneutrino dark matter in the extensions of MSSM\cite{ndm1,ndm2,ndm3,ndm4,ndm5,ndm6,ndm7,ndm8}.

After this introduction, we show the main contents of BLMSSM in section 2. Sections 3 and 4 are devoted to the formulas of relic density and direct detection.
We calculate the numerical results and find reasonable parameter space in BLMSSM in section 5. The discussion and conclusion are shown in section 6.

\section{the BLMSSM}
The detection of the lightest CP-even Higgs at LHC\cite{exp1,exp2,exp3} has proved that the SM achieves great success.
 Extending the MSSM with the local gauge group $SU(3)_C\times SU(2)_L \times U(1)_Y\times U(1)_{B}\times U(1)_{L}$,
 physicists obtain the BLMSSM\cite{BL1,BL2}. The exotic leptons and  exotic quarks are respectively introduced for canceling L and B anomalies. The superfields in BLMSSM are shown in TABLE.I.
 The Higgs superfields(two doublets and four singlets) obtain nonzero vacuum expectation values (VEVs).
So they break both lepton number and baryon number spontaneously.

After the Higgs obtain VEVs, the local gauge symmetry $SU(2)_{L}\otimes U(1)_{Y}\otimes U(1)_{B}\otimes U(1)_{L}$ breaks down to the electromagnetic symmetry $U(1)_{e}$. If we mark the nonzero VEVs of the $SU(2)_L$ singlets $\Phi_{B},\;\varphi_{B},\;\Phi_{L},\;\varphi_{L}$ and the $SU(2)_L$ doublets $H_{u},\;H_{d}$ as $\upsilon_{{B}},\;\overline{\upsilon}_{{B}},\;\upsilon_{L},\;\overline{\upsilon}_{L},\;\upsilon_{u}$ and $\upsilon_{d}$, we have
\begin{eqnarray}
&&H_{u}=\left(\begin{array}{c}H_{u}^+\\{1\over\sqrt{2}}\Big(\upsilon_{u}+H_{u}^0+iP_{u}^0\Big)\end{array}\right)\;,~~~~
H_{d}=\left(\begin{array}{c}{1\over\sqrt{2}}\Big(\upsilon_{d}+H_{d}^0+iP_{d}^0\Big)\\H_{d}^-\end{array}\right)\;,
\nonumber\\
&&\Phi_{B}={1\over\sqrt{2}}\Big(\upsilon_{B}+\Phi_{B}^0+iP_{B}^0\Big)\;,~~~~~~~~~
\varphi_{B}={1\over\sqrt{2}}\Big(\overline{\upsilon}_{B}+\varphi_{B}^0+i\overline{P}_{B}^0\Big)\;,
\nonumber\\
&&\Phi_{L}={1\over\sqrt{2}}\Big(\upsilon_{L}+\Phi_{L}^0+iP_{L}^0\Big)\;,~~~~~~~~~~
\varphi_{L}={1\over\sqrt{2}}\Big(\overline{\upsilon}_{L}+\varphi_{L}^0+i\overline{P}_{L}^0\Big)\;.
\label{VEVs}
\end{eqnarray}

\begin{table}
\begin{center}
\caption[]{The superfields in the BLMSSM.}
\begin{tabular*}{87.0mm}{|c|c|c|c|c|c|}
\hline
superfields &$SU(3)_C$ &$SU(2)_L$ &$U(1)_Y$ &$U(1)_B$ &$U(1)_L$ \\
\hline
$\hat{Q}$&3 &2 &1/6 &1/3 &0 \\
\hline
$\hat{U}^c$&$\bar{3}$ &1 &-2/3 &-1/3 &0 \\
\hline
$\hat{D}^c$&$\bar{3}$ &1 &1/3 &-1/3 &0 \\
\hline
$\hat{L}$&1 &2 &-1/2 &0 &1 \\
\hline
$\hat{E}^c$&1 &1 &1 &0 &-1 \\
\hline
$\hat{N}^c$&1 &1 &0 &0 &-1 \\
\hline
$\hat{Q}_4$&3 &2 &1/6 &$B_4$ &0 \\
\hline
$\hat{U}_4^c$&$\bar{3}$ &1 &-2/3 &-$B_4$ &0 \\
\hline
$\hat{D}_4^c$&$\bar{3}$ &1 &1/3 &-$B_4$ &0 \\
\hline
$\hat{L}_4$&1 &2 &-1/2 &0 &$L_4$ \\
\hline
$\hat{E}_4^c$&1 &1 &1 &0 &-$L_4$ \\
\hline
$\hat{N}_4^c$&1 &1 &0 &0 &-$L_4$ \\
\hline
$\hat{Q}_5^c$&$\bar{3}$ &2 &-1/6 &-1-$B_4$ &0 \\
\hline
$\hat{U}_5$&3 &1 &2/3 &1+$B_4$ &0 \\
\hline
$\hat{D}_5$&3 &1 &-1/3 &1+$B_4$ &0 \\
\hline
$\hat{L}_5^c$&1 &2 &1/2 &0 &-3-$L_4$ \\
\hline
$\hat{E}_5$&1 &1 &-1 &0 &3+$L_4$ \\
\hline
$\hat{N}_5$&1 &1 &0 &0 &3+$L_4$ \\
\hline
$\hat{H}_u$&1 &2 &1/2 &0 &0 \\
\hline
$\hat{H}_d$&1 &2 &-1/2 &0 &0 \\
\hline
$\hat{\Phi}_B$&1 &1 &0 &1 &0 \\
\hline
$\hat{\varphi}_B$&1 &1 &0 &-1 &0 \\
\hline
$\hat{\Phi}_L$&1 &1 &0 &0 &-2 \\
\hline
$\hat{\varphi}_L$&1 &1 &0 &0 &2 \\
\hline
$\hat{X}$&1 &1 &0 &2/3+$B_4$ &0 \\
\hline
$\hat{X}'$&1 &1 &0 &-2/3-$B_4$ &0 \\
\hline
\end{tabular*}%
\end{center}
\end{table}

The superpotential of BLMSSM is\cite{BL3}
\begin{eqnarray}
&&{\cal W}_{{BLMSSM}}={\cal W}_{{MSSM}}+{\cal W}_{B}+{\cal W}_{L}+{\cal W}_{X}\;,
\label{superpotential1}
\nonumber\\&&{\cal W}_{B}=\lambda_{Q}\hat{Q}_{4}\hat{Q}_{5}^c\hat{\Phi}_{B}+\lambda_{U}\hat{U}_{4}^c\hat{U}_{5}
\hat{\varphi}_{B}+\lambda_{D}\hat{D}_{4}^c\hat{D}_{5}\hat{\varphi}_{B}+\mu_{B}\hat{\Phi}_{B}\hat{\varphi}_{B}
\nonumber\\
&&\hspace{1.2cm}
+Y_{{u_4}}\hat{Q}_{4}\hat{H}_{u}\hat{U}_{4}^c+Y_{{d_4}}\hat{Q}_{4}\hat{H}_{d}\hat{D}_{4}^c
+Y_{{u_5}}\hat{Q}_{5}^c\hat{H}_{d}\hat{U}_{5}+Y_{{d_5}}\hat{Q}_{5}^c\hat{H}_{u}\hat{D}_{5}\;,
\nonumber\\
&&{\cal W}_{L}=Y_{{e_4}}\hat{L}_{4}\hat{H}_{d}\hat{E}_{4}^c+Y_{{\nu_4}}\hat{L}_{4}\hat{H}_{u}\hat{N}_{4}^c
+Y_{{e_5}}\hat{L}_{5}^c\hat{H}_{u}\hat{E}_{5}+Y_{{\nu_5}}\hat{L}_{5}^c\hat{H}_{d}\hat{N}_{5}
\nonumber\\
&&\hspace{1.2cm}
+Y_{\nu}\hat{L}\hat{H}_{u}\hat{N}^c+\lambda_{{N^c}}\hat{N}^c\hat{N}^c\hat{\varphi}_{L}
+\mu_{L}\hat{\Phi}_{L}\hat{\varphi}_{L}\;,
\nonumber\\
&&{\cal W}_{X}=\lambda_1\hat{Q}\hat{Q}_{5}^c\hat{X}+\lambda_2\hat{U}^c\hat{U}_{5}\hat{X}^\prime
+\lambda_3\hat{D}^c\hat{D}_{5}\hat{X}^\prime+\mu_{X}\hat{X}\hat{X}^\prime\;.
\label{superpotential-BL}
\end{eqnarray}
where ${\cal W}_{{MSSM}}$ is the superpotential of the MSSM. The soft breaking terms $\mathcal{L}_{{soft}}$ of the BLMSSM can be written as\cite{BL3}.
\begin{eqnarray}
&&{\cal L}_{{soft}}={\cal L}_{{soft}}^{MSSM}-(m_{{\tilde{N}^c}}^2)_{{IJ}}\tilde{N}_I^{c*}\tilde{N}_J^c
-m_{{\tilde{Q}_4}}^2\tilde{Q}_{4}^\dagger\tilde{Q}_{4}-m_{{\tilde{U}_4}}^2\tilde{U}_{4}^{c*}\tilde{U}_{4}^c
-m_{{\tilde{D}_4}}^2\tilde{D}_{4}^{c*}\tilde{D}_{4}^c
\nonumber\\
&&\hspace{1.3cm}
-m_{{\tilde{Q}_5}}^2\tilde{Q}_{5}^{c\dagger}\tilde{Q}_{5}^c-m_{{\tilde{U}_5}}^2\tilde{U}_{5}^*\tilde{U}_{5}
-m_{{\tilde{D}_5}}^2\tilde{D}_{5}^*\tilde{D}_{5}-m_{{\tilde{L}_4}}^2\tilde{L}_{4}^\dagger\tilde{L}_{4}
-m_{{\tilde{\nu}_4}}^2\tilde{N}_{4}^{c*}\tilde{N}_{4}^c
\nonumber\\
&&\hspace{1.3cm}
-m_{{\tilde{e}_4}}^2\tilde{E}_{_4}^{c*}\tilde{E}_{4}^c-m_{{\tilde{L}_5}}^2\tilde{L}_{5}^{c\dagger}\tilde{L}_{5}^c
-m_{{\tilde{\nu}_5}}^2\tilde{N}_{5}^*\tilde{N}_{5}-m_{{\tilde{e}_5}}^2\tilde{E}_{5}^*\tilde{E}_{5}
-m_{{\Phi_{B}}}^2\Phi_{B}^*\Phi_{B}
\nonumber\\
&&\hspace{1.3cm}
-m_{{\varphi_{B}}}^2\varphi_{B}^*\varphi_{B}-m_{{\Phi_{L}}}^2\Phi_{L}^*\Phi_{L}
-m_{{\varphi_{L}}}^2\varphi_{L}^*\varphi_{L}-\Big(m_{B}\lambda_{B}\lambda_{B}
+m_{L}\lambda_{L}\lambda_{L}+h.c.\Big)
\nonumber\\
&&\hspace{1.3cm}
+\Big\{A_{{u_4}}Y_{{u_4}}\tilde{Q}_{4}H_{u}\tilde{U}_{4}^c+A_{{d_4}}Y_{{d_4}}\tilde{Q}_{4}H_{d}\tilde{D}_{4}^c
+A_{{u_5}}Y_{{u_5}}\tilde{Q}_{5}^cH_{d}\tilde{U}_{5}+A_{{d_5}}Y_{{d_5}}\tilde{Q}_{5}^cH_{u}\tilde{D}_{5}
\nonumber\\
&&\hspace{1.3cm}
+A_{{BQ}}\lambda_{Q}\tilde{Q}_{4}\tilde{Q}_{5}^c\Phi_{B}+A_{{BU}}\lambda_{U}\tilde{U}_{4}^c\tilde{U}_{5}\varphi_{B}
+A_{{BD}}\lambda_{D}\tilde{D}_{4}^c\tilde{D}_{5}\varphi_{B}+B_{B}\mu_{B}\Phi_{B}\varphi_{B}
+h.c.\Big\}
\nonumber\\
&&\hspace{1.3cm}
+\Big\{A_{{e_4}}Y_{{e_4}}\tilde{L}_{4}H_{d}\tilde{E}_{4}^c+A_{{\nu_4}}Y_{{\nu_4}}\tilde{L}_{4}H_{u}\tilde{N}_{4}^c
+A_{{e_5}}Y_{{e_5}}\tilde{L}_{5}^cH_{u}\tilde{E}_{5}+A_{{\nu_5}}Y_{{\nu_5}}\tilde{L}_{5}^cH_{d}\tilde{N}_{5}
\nonumber\\
&&\hspace{1.3cm}
+A_{N}Y_{\nu}\tilde{L}H_{u}\tilde{N}^c+A_{{N^c}}\lambda_{{N^c}}\tilde{N}^c\tilde{N}^c\varphi_{L}
+B_{L}\mu_{L}\Phi_{L}\varphi_{L}+h.c.\Big\}
\nonumber\\
&&\hspace{1.3cm}
+\Big\{A_1\lambda_1\tilde{Q}\tilde{Q}_{5}^cX+A_2\lambda_2\tilde{U}^c\tilde{U}_{5}X^\prime
+A_3\lambda_3\tilde{D}^c\tilde{D}_{5}X^\prime+B_{X}\mu_{X}XX^\prime+h.c.\Big\}\;.
\label{soft-breaking}
\end{eqnarray}
${\cal L}_{{soft}}^{MSSM}$ denote the soft breaking terms of the MSSM.

The elements of the mass squared matrix of sneutrino read as\cite{BL4}
\begin{eqnarray}
&&M_{\tilde n}^2(\tilde\nu_I^* \tilde\nu_J)=\frac{g_1^2+g_2^2}{8}(v_d^2-v_u^2)\delta_{IJ}+g_L^2(\bar{v}_L^2-v_L^2)\delta_{IJ}+\frac{v_u^2}{2}(Y_\nu^{\dag} Y_\nu)_{IJ}+(m_{\tilde L}^2)_{IJ},
\nonumber\\
&&M_{\tilde n}^2(\tilde N_I ^{c*} \tilde N _J ^c)=-g_L^2(\bar{v}_L^2-v_L^2)\delta_{IJ}+\frac{v_u^2}{2}(Y_\nu^{\dag} Y_\nu)_{IJ}+2\bar{v}_L^2({\lambda}_{N^c}^{\dag} {\lambda}_{N^c})_{IJ}
+(m_{\tilde N^c} ^2)_{IJ}\nonumber\\&&\hspace{2.0cm}
+{\mu}_L\frac{v_L}{\sqrt 2}(\lambda_{N^c})_{IJ}-\frac{\bar v _L}{\sqrt 2}(A_{N^c})_{IJ}({\lambda}_{N^c})_{IJ},
\nonumber\\&&
M_{\tilde n}^2(\tilde\nu_I \tilde N_J^c)={\mu}^* \frac{v_d}{\sqrt 2}(Y_\nu)_{IJ}-v_u \bar v_L(Y_\nu^{\dag}\lambda_{N^c})_{IJ}+\frac{v_u}{\sqrt 2}(A_N)_{IJ}(Y_\nu)_{IJ}.
\end{eqnarray}
The scalar neutrino mass squared matrix is diagonalized by the matrix $Z_{\tilde{\nu}}${\cite{matrix}}.
The lightest mass eigenstate of scalar neutrino is supposed as dark matter in this work.

With the introduced  superfields $\hat{N}^c$, three neutrinos obtain tiny masses through the see-saw mechanism.
The mass matrix of neutrinos is shown in the basis $(\psi_{\nu_L^I}, \psi_{N_R^{cI}})$,
\begin{eqnarray}
&&Z_{N_{\nu}}^\top\left(\begin{array}{cc}
  0&\frac{v_u}{\sqrt{2}}(Y_{\nu})^{IJ} \\
   \frac{v_u}{\sqrt{2}}(Y^{T}_{\nu})^{IJ}  & \frac{\bar{v}_L}{\sqrt{2}}(\lambda_{N^c})^{IJ}
    \end{array}\right)Z_{N_{\nu}}=diag(m_{\nu^{\alpha}}), \alpha=1\cdot\cdot\cdot6,I,J=1,2,3,
\nonumber\\&& \psi_{{\nu_L^I}}=Z_{{N_{\nu}}}^{I\alpha}k_{N_\alpha}^0,\;\;\;\;
\psi_{N_R^{cI}}=Z_{{N_{\nu}}}^{(I+3)\alpha}k_{N_\alpha}^0,\;\;\;\;
\chi_{N_\alpha}^0=\left(\begin{array}{c}
   k_{N_\alpha}^0\\  \bar{k}_{N_\alpha}^0
    \end{array}\right),
\end{eqnarray}
with $\chi_{N_\alpha}^0$ representing the mass eigenstates of neutrino fields.

In BLMSSM, the mass squared matrix of the slepton reads as
\begin{eqnarray}
\left(\begin{array}{cc}({\cal M}^2_L)_{LL}&({\cal M}^2_L)_{LR}\\({\cal M}^2_L)^\dag_{LR}&({\cal M}^2_L)_{RR}\end{array}\right).
\end{eqnarray}
${({\cal M}^2_L)}_{LL}$, ${(\cal M}^2_L)_{LR}$ and ${({\cal M}^2_L)}_{RR}$ are shown here
\begin{eqnarray}
&&({\cal M}^2_L)_{LL}=\frac{(g^2_1-g^2_2)(v^2_d-v^2_u)}{8} \delta_{IJ}+g^2_L(\bar{v}^2_L-v^2_L)\delta_{IJ}+{m^2_{l^I}}\delta_{IJ}+(m^2_{\bar{L}})_{IJ},
\nonumber\\
&&({\cal M}^2_L)_{LR}={\mu^\ast v_u\over\sqrt{2}}(Y_l)_{IJ}-{v_u\over\sqrt{2}}(A'_l)_{IJ}+{v_d\over\sqrt{2}}(A_l)_{IJ},
\nonumber\\
&&({\cal M}^2_L)_{RR}={g^2_1(v^2_u-v^2_d)\over4}\delta_{IJ}-g^2_L(\bar{v}^2_L-v^2_L)\delta_{IJ}+m^2_{l^I}\delta_{IJ}+(m^2_{\tilde{R}})_{IJ}.
\end{eqnarray}
We rotate this mass squared matrix to the mass eigenstates through the unitary matrix $Z_{\tilde{L}}$.

Some couplings are shown here. The couplings of W-lepton-neutrino and Z-neutrino-neutrino are different from those in MSSM, and their
concrete forms are
\begin{eqnarray}
&&\mathcal{L}_{Wl\nu}=-\frac{e}{\sqrt{2}s_W}W_{\mu}^+\sum_{I=1}^3\sum_{\alpha=1}^6Z_{N_{\nu}}^{I\alpha*}\bar{\chi}_{N_{\alpha}}^0\gamma^{\mu}P_Ll^I+h.c.,
\nonumber\\&&\mathcal{L}_{Z\nu\nu}=-\frac{e}{2s_Wc_W}Z_{\mu}\sum_{I=1}^3\sum_{\alpha,\beta=1}^6
Z_{N_{\nu}}^{I\alpha*}Z_{N_{\nu}}^{I\beta}\bar{\chi}_{N_{\alpha}}^0\gamma^{\mu}P_L\chi_{N_{\beta}}^0+h.c.,
\end{eqnarray}
where $s_W(c_W)$ represents $\sin\theta_W(\cos\theta_W)$. $\theta_W$ is the Weinberg angle.

The Z-sneutrino-sneutrino coupling is deduced as:
\begin{eqnarray}
&&\mathcal{L}_{Z\tilde{\nu}\tilde{\nu}}=-\frac{e}{2s_Wc_W}Z_{\mu}\sum_{I=1}^3\sum_{i,j=1}^6Z_{\tilde{\nu}}^{Ii*}Z_{\tilde{\nu}}^{Ij}\tilde{\nu}^{i*}i(\overrightarrow{\partial}^{\mu}
-\overleftarrow{\partial}^{\mu})\tilde{\nu}^j.
\end{eqnarray}

We also obtain the chargino-lepton-sneutrino and neutralino-neutrino-sneutrino couplings
\begin{eqnarray}
&&\mathcal{L}_{\chi^{\pm}l\tilde{\nu}}=-\sum_{I=1}^3\sum_{i=1}^6\sum_{j=1}^2\bar{\chi}^-_j
\Big(Y_l^{I} Z_-^{2j*}Z_{\tilde{\nu}}^{Ii*}P_R+
[\frac{e}{s_W}Z_+^{1j}Z_{\tilde{\nu}}^{Ii*}
\nonumber\\&&\hspace{1.8cm}+Y_\nu^{Ii}Z_+^{2j}Z_{\tilde{\nu}}^{(I+3)i*}]P_L
\Big)l^I\tilde{\nu}^{i*}+h.c.
\nonumber\\
&&{\cal L}_{\chi^0 \nu\tilde{\nu}}=\sum_{I,J=1}^3\sum_{i=1}^4\sum_{\alpha,k=1}^6\bar{\chi}^0_i
\Big(\frac{e}{\sqrt{2}s_Wc_W}(Z_N^{1i}s_W-Z_N^{2i}c_W)Z_{\tilde{\nu}}^{Ik*}Z_{N_{\nu}}^{I\alpha}\nonumber\\&&+\frac{Y_\nu^{IJ}}{\sqrt{2}}Z_N^{4i}(Z_{N_{\nu}}^{I\alpha}
Z_{\tilde{\nu}}^{(J+3)k*}+Z_{N_{\nu}}^{(I+3)\alpha}
Z_{\tilde{\nu}}^{Jk*})\Big)P_L\chi^{0}_{N_\alpha}\tilde{\nu}^{k*}+h.c..
\end{eqnarray}
Here, $Z_-$ and $Z_+$ are used to diagonalize chargino mass matrix.

The charged Higgs-lepton-neutrino couplings read as
\begin{eqnarray}
&&{\cal L}_{H^\pm L\nu}=\sum^3_{I,J=1}\sum^6_{\alpha=1}G^\pm\bar{e}^J(Y^{I}_l\cos\beta Z_{N_{\nu}}^{I\alpha}\delta_{IJ}
P_L-Y^{IJ\ast}_\nu\sin\beta Z_{N_{\nu}}^{(I+3) \alpha}P_R)\chi^0_{N_\alpha}
\nonumber\\&&\hspace{1.5cm}
-\sum^3_{I,J=1}\sum^6_{\alpha=1}H^\pm\bar{e}^J(Y^{I}_l\sin\beta Z_{N_{\nu}}^{I\alpha}\delta_{IJ}P_L+Y^{IJ\ast}_\nu\cos\beta Z_{N_{\nu}}^{(I+3)\alpha}P_R)\chi^0_{N_\alpha}+h.c.
\end{eqnarray}

The coupling of CP-even Higgs with sneutrinos are
   \begin{eqnarray}
  &&\mathcal{L}_{\tilde{\nu}\tilde{\nu}H^0}=\sum_{i,j=1}^6\tilde{N}^{i*}\tilde{N}^{j}\Big(H^0[(N^u_M)_{ij}\sin\alpha+(N^d_M)_{ij}\cos\alpha]
  \nonumber\\&&\hspace{1.7cm}+h^0[(N^u_M)_{ij}\cos\alpha-(N^d_M)_{ij}\sin\alpha]\Big),
\nonumber\\
  &&(N^u_M)_{ij}=\sum_{I=1}^3\Big(\frac{e^2}{4s_W^2c_W^2}v_u Z_{\tilde{\nu}}^{Ii*}Z_{\tilde{\nu}}^{Ij}-\sum_{J=1}^3v_u|Y_{\nu}^{IJ}|^2\delta_{ij}
  +(\lambda_{\nu_c}^*\bar{v}_L-\frac{A_N}{\sqrt{2}})Z_{\tilde{\nu}}^{(I+3)i*}Z_{\tilde{\nu}}^{Ij}\Big),\nonumber\\&&
  (N^d_M)_{ij}=-\sum_{I=1}^3\frac{e^2}{4s_W^2c_W^2}v_d Z_{\tilde{\nu}}^{Ii*}Z_{\tilde{\nu}}^{Ij}
  -\sum_{I,J=1}^3\frac{\mu^*}{\sqrt{2}}Y_{\nu}^{IJ}Z_{\tilde{\nu}}^{(I+3)i*}Z_{\tilde{\nu}}^{Jj}.
  \end{eqnarray}

\section{relic density}
We suppose the lightest sneutrino $(\tilde{\nu}_j)$ as a dark matter candidate in this paper and its main element is right-handed. Any dark matter candidate should satisfy the relic density constraint. Therefore,
the result of theoretical calculation the closer to the actual observed value, the more reliable this theory becomes. The $\tilde{\nu}_j$ number density $n_{\tilde{\nu}_j}$ is given by the Boltzmann  equation\cite{dm2,dm5,dm6,dm7}
\begin{eqnarray}
\frac{d n_{\tilde{\nu}_j}}{dt}=-3Hn_{\tilde{\nu}_j}-\langle\sigma v\rangle_{SA}(n^2_{\tilde{\nu}_j}-n^2_{\tilde{\nu}_j eq})
-\langle\sigma v\rangle_{CA}(n_{\tilde{\nu}_j}n_\phi-n_{\tilde{\nu}_j eq}n_{\phi eq}).
\end{eqnarray}
$\tilde{\nu}_j$ can both self-annihilate and co-annihilate with another field of particle $\phi$. When the annihilation rate of $\tilde\nu_{j}$ becomes roughly equal to the Hubble expansion rate, the species freeze out at the temperature $T_F$,
\begin{eqnarray}
\langle\sigma v\rangle_{SA}n_{\tilde{\nu}_j}+\langle\sigma v\rangle_{CA}n_{\phi}\sim H(T_F).
\end{eqnarray}
If we suppose $M_{\phi}>M_{\tilde\nu_{j}}$\cite{ndm9}
\begin{eqnarray}
n_\phi=\Big(\frac{M_\phi}{M_{\tilde{\nu}_j}}\Big)^{3/2}\texttt{Exp}[(M_{\tilde{\nu}_j}-M_\phi)/T]n_{\tilde{\nu}_j}.
\end{eqnarray}
Then it becomes
\begin{eqnarray}
\Big[\langle\sigma v\rangle_{SA}+\langle\sigma v\rangle_{CA}
\Big(\frac{M_\phi}{M_{\tilde{\nu}_j}}\Big)^{3/2}\texttt{Exp}[(M_{\tilde{\nu}_j}-M_\phi)/T]\Big]n_{\tilde{\nu}_j}\sim H(T_F).
\end{eqnarray}

We calculate the self-annihilation cross section $\sigma(\tilde{\nu}_j \tilde{\nu}_j^* \rightarrow$ anything) and
co-annihilation cross section $\sigma(\tilde{\nu}_j \phi \rightarrow$ anything)
to study its annihilation rate $\langle\sigma v\rangle_{SA}$ ($\langle\sigma v\rangle_{CA}$) and its relic density $\Omega_D$ in the thermal history of the universe.
In the frame of the central mass system, the annihilation rate can be written as $\sigma v=a+bv^2$\cite{dm2,dm7,dm8,dm9}, where $v$ represents the relative velocity of the two particles in the initial states. The thermally averaged cross section times velocity $\langle \sigma_{eff} v\rangle$ was extended by the Edsj\"{o} and
Gondolo, which includes the co-annihilations \cite{exten1,exten2}
\begin{eqnarray}
\langle \sigma_{eff} v\rangle(x)
=\frac{\int_2^\infty K_1(\frac{a}{x})\sum_{i,j=1}^N\lambda(a^2,b_i^2,b_j^2)g_ig_j\sigma_{ij}(a)da}{4x\Big(\sum_{i=1}^NK_2(\frac{b_i}{x})b_i^2g_i\Big)^2},
\end{eqnarray}
with $x=\frac{T}{M_{\tilde{\nu}_j}}, ~\lambda(a^2,b_i^2,b_j^2)=a^4+b_i^4+b_j^4-2(a^2b_i^2+a^2b_j^2+b_i^2b_j^2)$, $a=\sqrt{s}/M_{\tilde{\nu}_j}$ and $b_i=m_i/M_{\tilde{\nu}_j}$.
 $g$ is the number of the relativistic degrees of freedom with mass less than $T_F$.
$\sigma_{ij}$ is the cross
section for the annihilation reaction $ij$ to any allowed final state including two SM and/or Higgs particles.
Then one can calculate the relic density, whose value is  $\Omega_D h^2=0.1186\pm 0.0020$\cite{pdg}.

\begin{figure}
  \setlength{\unitlength}{1mm}
  \centering
  \includegraphics[width=3.5in]{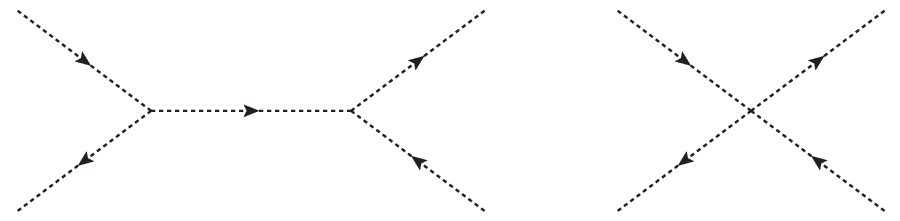}
  \caption[]{The Feynman diagrams of $\tilde{\nu_j} \tilde{\nu_j} \rightarrow h^0 h^0$}\label{pica}
\end{figure}

\begin{figure}
  \setlength{\unitlength}{1mm}
  \centering
  \includegraphics[width=4.0in]{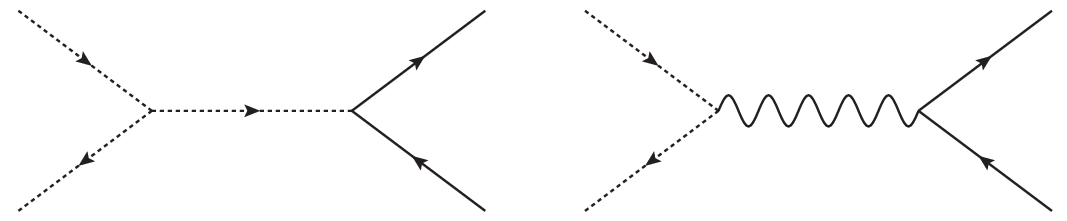}
  \caption[]{The Feynman diagrams of $\tilde \nu_j \tilde \nu_j \rightarrow \bar{q}q$}\label{pice}
\end{figure}

Ignoring some of the lower contributions, the main self-annihilation processes are: $\tilde{\nu}_j+\tilde{\nu}_j\rightarrow \{(W+W),~(Z+Z),~(h^0+h^0),~(\bar{l}_i+ l_i),~(\bar{\nu}_i+ \nu_i),~(\bar{u}_i+ u_i),~(\bar{d}_i+ d_i)\}$ with $i=1,~2,~3$, $h^0$ representing the lightest CP-even Higgs boson.
The self-annihilation processes are plotted in the FIG.\ref{pica}-FIG.\ref{pice}.

The studied co-annihilation processes read as\cite{SNzhao}:

a. $\tilde{\nu}_j+\tilde{\nu}_k\rightarrow \{(W+W),~(Z+Z),~(h^0+h^0),~(\bar{l}_i+ l_i)),~(\bar{\nu}_i+ \nu_i),~(\bar{u}_i+ u_i),~(\bar{d}_i+ d_i)\}$ with $k=2\dots6,~i=1,~2,~3$.

b. $\tilde{\nu}_j+\chi^0_k\rightarrow \{(W^++l^-_i),~(W^-+l^+_i),~(Z+\nu_i)\} $ and $k=1\dots4,~i=1\dots3$.

c. $\tilde{\nu}_j+\chi^\pm_k\rightarrow \{(l^{\pm}_i+\gamma),~(l^{\pm}_i+Z),~(l^{\pm}_i+h^0),~(W^{\pm}+\nu_i)\} $ and $k=1\dots2,~i=1\dots3$.

d. $\tilde{\nu}_j+\tilde{L}^\pm_k\rightarrow \{(W^{\pm}_i+Z),~(W^{\pm}_i+\gamma),~(W^{\pm}_i+h^0),~(\nu_i+l_i^\pm),~(u_i+\bar{d}_i),~(d_i+\bar{u}_i)\} $ and $k=1\dots6,~i=1\dots3$.

\begin{figure}
  \setlength{\unitlength}{1mm}
  \centering
  \includegraphics[width=3.6in]{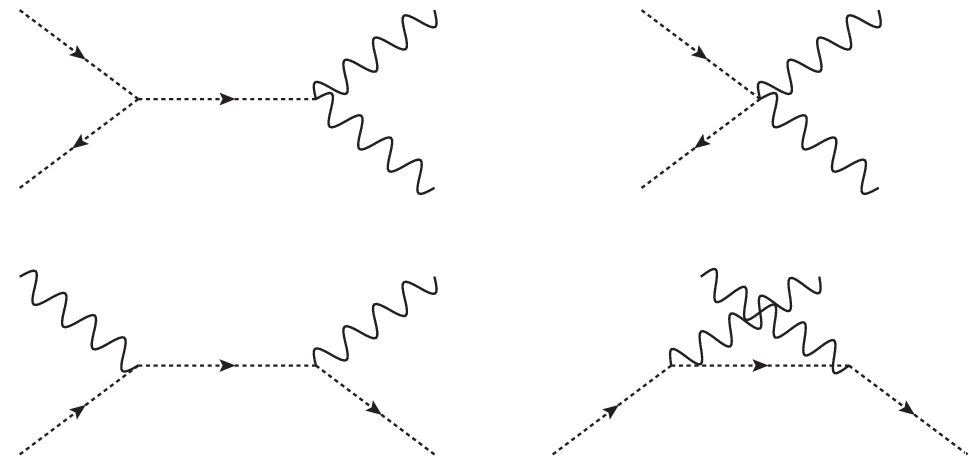}
  \caption[]{The Feynman diagrams of $\tilde \nu_j \tilde \nu_j \rightarrow ZZ$}\label{picb}
\end{figure}
\begin{figure}
  \setlength{\unitlength}{1mm}
  \centering
  \includegraphics[width=4.0in]{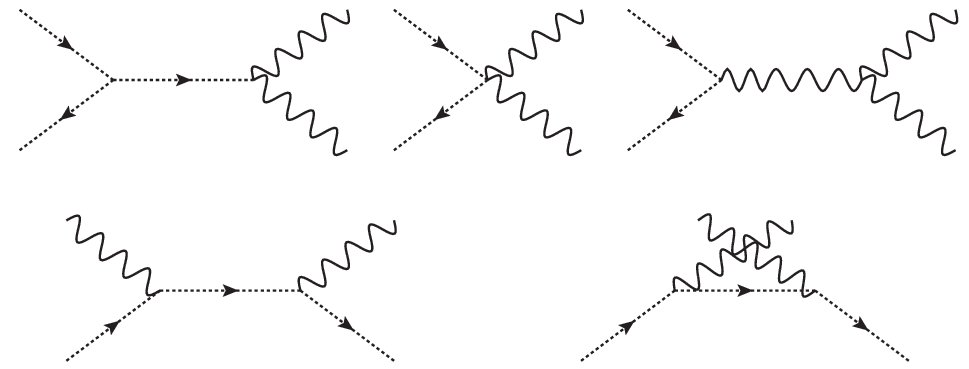}
  \caption[]{The Feynman diagrams of $\tilde \nu_j \tilde \nu_j \rightarrow WW$}\label{picc}
\end{figure}
\begin{figure}
  \setlength{\unitlength}{1mm}
  \centering
  \includegraphics[width=3.8in]{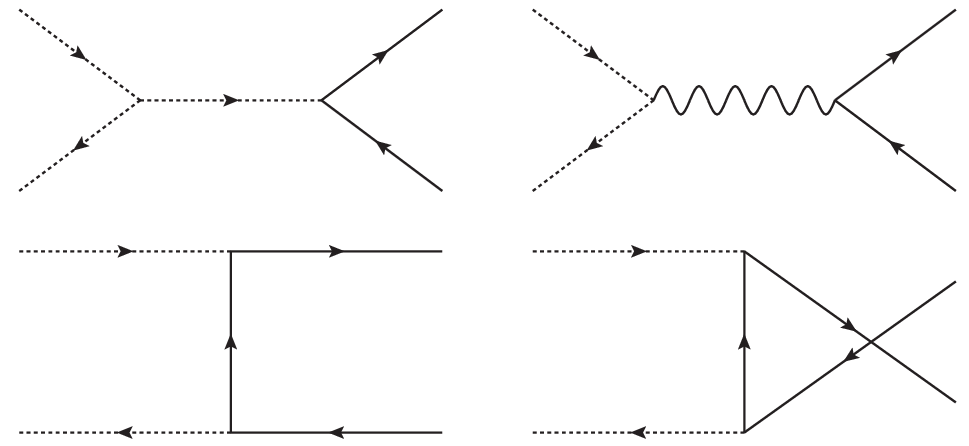}
  \caption[]{The Feynman diagrams of $\tilde \nu_j \tilde \nu_j \rightarrow \bar{\nu} \nu (\bar{l}l)$}\label{picd}
\end{figure}

For example, we show some results for the self-annihilation decays $\tilde \nu_j \tilde \nu_j\rightarrow h^0 h^0$ and $\tilde \nu_j \tilde \nu_j\rightarrow \bar{t}t$.
The Feynman diagrams for the process $\tilde \nu_j \tilde \nu_j\rightarrow h^0 h^0$ are plotted in the FIG.\ref{pica}.  The lightest mass eigenstate of CP-even Higgs $H^0_i (i=1,~2)$ is $H_1^0$ represented by $h^0$. The analytic results are
deduced here.
\begin{eqnarray}
&&\langle{\sigma v}\rangle_{h^0 h^0}=\frac{|M|^2}{64\pi M_{\tilde {\nu}_j}^2}\sqrt{1-\frac{m_{h^0}^2}{M_{\tilde {\nu}_j}^2}},\nonumber\\&&
{|M|}^2={|\mathcal{D}|}^2+{\sum_{i,k=1}^2\frac{\mathcal{B}_i \mathcal{B}_k^* \mathcal{C}_i \mathcal{C}_k^*}{(M_{\tilde {\nu}_j}^2-m_{H_i^0}^2)
(M_{\tilde {\nu}_j}^2-m_{H_k^0}^2)}}+2\Re[\sum_{i=1}^2\frac{\mathcal{D}^* \mathcal{B}_i\mathcal{C}_i}{(M_{\tilde {\nu}_j}^2 -m_{H_i^0}^2)}].
\end{eqnarray}
with
\begin{eqnarray}
&&\mathcal{B}_i=-\frac {e^2({A_R^{11} B_R^i +2A_R^{1i} B_R^1})}{4s_W^2 c_W^2},~~
\mathcal{C}_i=-\sum_1^3\frac{e^2 B_R^i{Z_{I1} {Z_{I1}^*}}}{4{s_W}^2 {c_W}^2},~~
\mathcal{D}=-\frac{e^2 A_R^{11}}{4s_W^2 c_W^2} \sum_{I=1}^3{Z_{I1} Z_{I1}^*},
\nonumber\\&&B_R^i=v_1 Z^{1i}_R-v_2 Z^{2i}_R,~~~~~~~~~~~~~
A_R^{ij}=Z_R^{1i} Z_R^{1j}-Z_R^{2i} Z_R^{2j}.
\end{eqnarray}
 $Z_R$ is the matrix to diagonalize the mass squared matrix of CP-even Higgs.

The $\langle{\sigma v}\rangle$ for the process $\tilde \nu_j \tilde \nu_j\rightarrow \bar{t}t$ is also shown here. This is the leading order
contribution from the virtual CP-even Higgs. The virtual Z boson contribution is suppressed by the square of the relative velocity and we do not show
it here.
\begin{eqnarray}
\langle{\sigma v}\rangle_{\bar t t}=\frac{3(M_{\tilde \nu_j}^2-{m_t^2})}{32 \pi M_{\tilde \nu_j}^2}
\frac{m^2_t}{v_u^2 }\sum_{i,k=1}^2\frac{8 \mathcal{C}_i \mathcal{C}_k^*}{(4M_{\tilde \nu_j}^2-m_{H_i^0}^2)(4M_{\tilde \nu_j}^2-{m_{H_k^0}^2})}.
\end{eqnarray}
\section{direct detection}
  At the quark level, the process for the direct detection is
$\tilde{\nu}+q\rightarrow \tilde{\nu}+q$.
The lightest sneutrino is dominantly composed of the right-handed sneutrino, which is not active.
Then it can easily satisfy the constraint from the direct detection experiments. This specialty is represented by the neutrino Yukawa coupling
in the mass squared matrix of sneutrino. It leads to the suppression factor $Z^{Ij}_{\nu}, ~I=(I=1,2,3)$ appearing in the Feynman rules for the coupling $Z-\tilde{\nu}-\tilde{\nu}$ and $H^0(A^0)-\tilde{\nu}-\tilde{\nu}$.  The lightest sneutrino is represented by $\tilde{\nu}_j$ and $j>3$, with the right-handed sneutrinos  labeled as 4, 5, 6. Then $Z^{Ij}_{\nu}$ is in direct proportion to
 the tiny neutrino Yukawa coupling $Y_\nu$. The order of $Y_\nu$ is in the region $(10^{-6}\sim10^{-9})$.

 The cross section from virtual Z boson contribution is
 suppressed by $Y_\nu^4$. While, the cross section of the process with virtual CP-odd Higgs boson  is
 suppressed by $Y_\nu^2 v^4$, with $v^4$ coming from $|\bar{q}\gamma_5 q|^2$\cite{LJfenxi}. The CP-even Higgs contribution is dominant with the factor $Y_\nu^2$.
 Therefore, the cross section for sneutrino scattering off nucleon
 is at least suppressed by $Y_\nu^2$. The large term is from the operator $\tilde{\nu}^{*}\tilde{\nu}\bar{q}q$ at quark level.
We should convert the quark level coupling to the effective nucleon coupling with the formulas\cite{LJfenxi}
\begin{eqnarray}
&&a_qm_q\bar{q}q\rightarrow f_Nm_N\bar{N}N,~~~~~~~~~~
f_N=\sum_{q=u,d,s}f_{Tq}^{(N)}a_q+\frac{2}{27}f_{TG}^{(N)}\sum_{q=c,b,t}a_q,\nonumber\\&&\langle N|m_q \bar{q}q|N\rangle=m_N f_{Tq}^{(N)},
~~~~~~~f_{TG}^{(N)}=1-\sum_{q=u,d,s}f_{Tq}^{(N)}.
\end{eqnarray}
The numbers of $f_{Tq}^{(N)}$ are \cite{FTshu,FTshu1,FTshu2},
\begin{eqnarray}
&&f^{(p)}_{Tu}=0.0153,~~~f^{(p)}_{Td}=0.0191,~~~f^{(p)}_{Ts}=0.0447, \nonumber\\&&
f^{(n)}_{Tu}=0.0110,~~~f^{(n)}_{Td}=0.0273,~~~f^{(n)}_{Ts}=0.0447.
\end{eqnarray}

With the obtained $f_N$, one gets the scattering cross section
\begin{eqnarray}
\sigma=\frac{1}{\pi}\mu_K^2[Z_pf_p+(A-Z_p)f_n]^2.
\end{eqnarray}
Here $\mu_K$ is the effective mass of the nucleon-sneutrino system,  $Z_p$ is the number of proton, and $A$ represents the number of atom.

\section{numerical results}
To confine the used parameter, we take into account the experiment results from the Particle Data Group(PDG)\cite{pdg}.
The results from our previous works are also considered\cite{SNzhao}.
The mass of the lightest neutral CP-even Higgs $h^0$ will take $m_{h^0}$=125.1 GeV\cite{exp1,exp2} in this work.

 The used  parameters are listed in the following
\begin{eqnarray}
&&\tan\beta=2,~ m_a=1{\rm TeV}, ~AN_{11}=AN_{22}=-450{\rm GeV},~AN_{33}=-40{\rm GeV},
\nonumber\\&&V_L=3{\rm TeV},~\tan\beta_L=2,~\lambda_{11} = \lambda_{22} =1,~\lambda_{33} = -0.1,~
M_1 = M_2 =1{\rm TeV},
\nonumber\\&&~\mu=0.8{\rm TeV},~g_L=\frac{1}{6},~ML_{11}=ML_{22}=ML_{33}=1{\rm TeV}^2,~\mu_L = 0.5{\rm TeV}.
\end{eqnarray}
\begin{figure}
  \setlength{\unitlength}{1mm}
  \centering
  \includegraphics[width=4in]{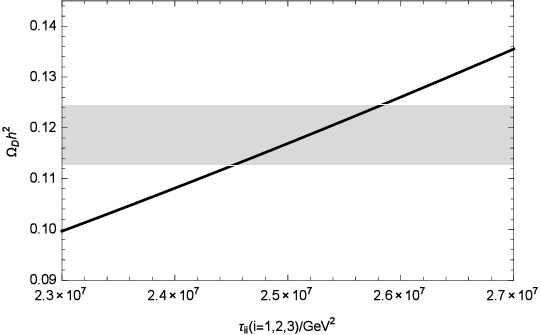}
  \caption[]{The relationship between $\Omega_D h^2$ and $\tau_{ii}$}\label{tau}
\end{figure}
\begin{figure}
  \setlength{\unitlength}{1mm}
  \centering
  \includegraphics[width=4in]{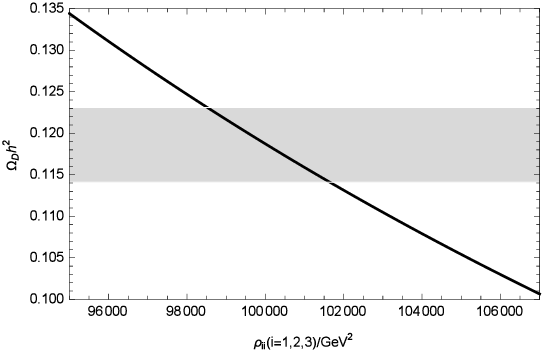}
  \caption[]{The relationship between $\Omega_D h^2$ and $\rho_{ii}$}\label{rho}
\end{figure}
Besides the fixed parameters shown above, there are some adjustable parameters defined as
\begin{eqnarray}
\tau_{ii}=(M^2_{\tilde{E}})_{ii},~~ \rho_{ii}=(M^2_{\tilde{\nu}})_{ii},~~\xi_{ii}=AL_{ii},~~\epsilon_{ii}=AL^\prime_{ii},~(i=1,~2,~3).
\end{eqnarray}
When we set their values as $\tau_{ii}=25.2{\rm TeV}^2,~\rho_{ii}=0.1{\rm TeV}^2,~\xi_{ii}=14{\rm TeV},~\epsilon_{ii}=3{\rm TeV},$
  and all the off-diagonal elements of the matrixes$((M^2_{\tilde{E}})_{ij}, (M^2_{\tilde{\nu}})_{ij},AL_{ij},AL^\prime_{ij}$ with $i\neq j)$
  are zero, the numerical result of dark matter relic density is  $\Omega_D h^2$=0.118581.
  It is very close to the experimental central value and in  one $\sigma$ sensitivity.

At this point, the lightest scalar neutrino mass is 350GeV, and the other scalar neutrinos' masses are all larger than 1TeV.
Besides, the lightest scalar lepton mass is around 1073GeV, and the masses of heavier scalar leptons are at several TeV order.
The heavy CP-even higgs mass is about 1TeV. The masses of neutralinos are in the region
$768\sim1033$ GeV. The two mass eigenstates of charginos are 1021GeV and 781GeV. Therefore, the lightest scalar neutrino is indeed
the LSP in this condition. Considering the masses of above particles, the resonance effect can not take place, because
the exchanged particle in S channel is not near $2\times350=700$GeV.

Then we study how these variables affect the results. $\tau_{ii}$ are the diagonal elements in the scalar lepton mass squared matrix, which
can influence the scalar lepton masses. Similarly, $\rho_{ii}$ are the diagonal elements of the scalar neutrino mass squared matrix.
The non-diagonal elements of the scalar lepton mass squared matrix include the parameters $\xi_{ii}$ and $\epsilon_{ii}$. In the whole,
$\tau_{ii}$, $\rho_{ii}$, $\xi_{ii}$ and $\epsilon_{ii}$ give effects to the scalar leptons (scalar neutrinos) masses and mixing. Therefore,
the relic density are influenced by these parameters.

In the FIG.\ref{tau}, we keep the parameters $\rho_{ii}=0.1{\rm TeV}^2,~\xi_{ii}=14{\rm TeV},~\epsilon_{ii}=3{\rm TeV},$
and plot the results of $\Omega_D h^2$ with $\tau_{ii}$.
The $\Omega_D h^2$ is an weakly increasing function, as $\tau_{ii}$ is on the specified
interval from 23 TeV$^2$ to 27 TeV$^2$. When $\tau_{ii}$ varies from 24.6 TeV$^2$ to 25.8 TeV$^2$,
the $\Omega_D h^2$ is in the band of $3 \sigma$ sensitivity denoted by the gray area in the FIG.\ref{tau}.
From these data, we can find that the masses of various particles are below a few TeV,
which is consistent with the constraints from LHC. So we believe that our results may be verified by the experiment before long.
Since the energy range of the particles is fairly wide, it is more advantageous to test them by experiments in the future.

On the contrary, the FIG.\ref{rho} shows that $\Omega_D h^2$ is the decreasing function of $\rho_{ii}$ within the selected interval (95000$\sim$107000)GeV$^2$. The value of $\Omega_D h^2$ decreases rapidly as the value of $\rho_{ii}$ increases. The reason is that the larger $\rho_{ii}$, the heavier scalar neutrino masses we obtain. At the same time, the matrix to diagonalize scalar neutrino mass matrix is changed, which influences the results directly and obviously. When  $\rho_{ii}$ is approximately from $9.85\times10^4 {\rm GeV^2}$ to $1.01\times10^5 {\rm GeV^2}$, we can ensure that $\Omega_D h^2$ is within a reasonable range of $3 \sigma$.

\begin{figure}
  \setlength{\unitlength}{1mm}
  \centering
  \includegraphics[width=4in]{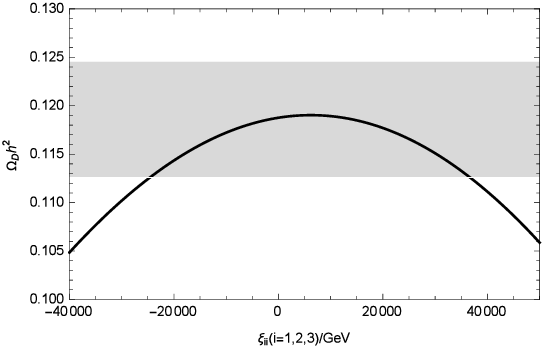}
  \caption[]{The relationship between $\Omega_D h^2$ and $\xi_{ii}$}\label{xi}
\end{figure}
\begin{figure}
  \setlength{\unitlength}{1mm}
  \centering
  \includegraphics[width=4in]{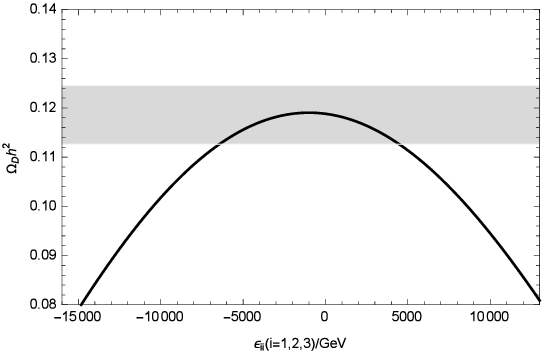}
  \caption[]{The relationship between $\Omega_D h^2$ and $\epsilon_{ii}$}\label{epsilon}
\end{figure}
The relationship of $\Omega_D h^2$ with $\xi_{ii}$ and $\epsilon_{ii}$ is a little bit more interesting. In the FIG.\ref{xi} and FIG.\ref{epsilon},
the values of $\Omega_D h^2$ are almost quadratic functions of the variables $\xi_{ii}$ and $\epsilon_{ii}$ respectively.
As $\xi_{ii}$ is in the range of -25 TeV to 35 TeV, the values of $\Omega_D h^2$ are all acceptable.
The behavior of $\Omega_D h^2$ versus $\epsilon_{ii}$ in the FIG.\ref{epsilon} is similar as the condition in the FIG.\ref{xi}.
To satisfy the experimental constraint, the region of $\epsilon_{ii}$ changing from -6 TeV to
5 TeV is acceptable. From the FIG.\ref{xi} and FIG.\ref{epsilon}, we can get the best values for $\xi_{ii}=14{\rm TeV}$ and $\epsilon_{ii}=3{\rm TeV}$.

\begin{figure}
  \setlength{\unitlength}{1mm}
  \centering
  \includegraphics[width=3.9in]{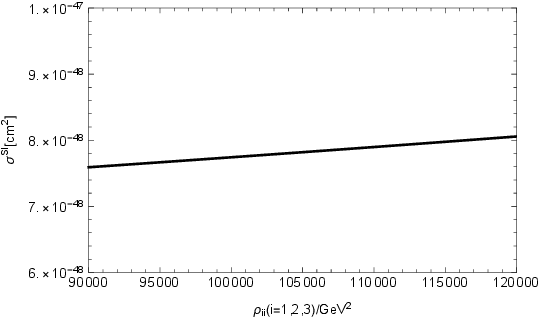}
  \caption[]{The relationship between the spin-independent cross section and $\epsilon_{ii}$}\label{HT1}
\end{figure}

With the parameters satisfying the relic density, which is shown in the front parts of this section,
we study the spin-independent cross section $\sigma^{SI}$ for sneutrino scattering off nucleon versus $\epsilon_{ii}$ in the FIG.\ref{HT1}.
$\sigma^{SI}$ increases slightly with the enlarging $\epsilon_{ii}$ from $9\times10^4$ GeV$^2$ to $12\times10^4$ GeV$^2$. The corresponding
theoretical value of $\sigma^{SI}$ is in the region $(7.8\sim8.05)\times 10^{-48}$cm$^2$.
 The lightest sneutrino mass is about 350 GeV, whose constraint from the direct detection experiments of $\sigma^{SI}$
 is around $3.0 \times 10^{-46}{\rm cm}^2$\cite{Hetan,Hetan1}. Our numerical results of the 
 spin-independent cross section is about two orders smaller than the experimental constraint.

 \begin{figure}
  \setlength{\unitlength}{1mm}
  \centering
  \includegraphics[width=3.9in]{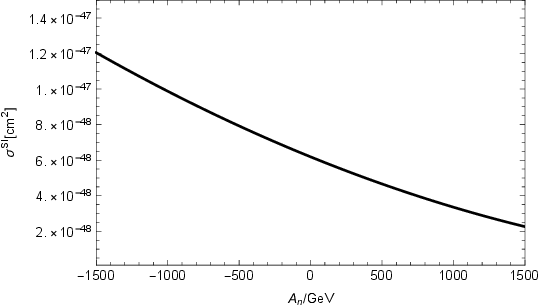}
  \caption[]{The relationship between the spin-independent cross section and $A_{n}$}\label{HT2}
\end{figure}

In the FIG.\ref{HT2}, we plot $\sigma^{SI}$ versus $A_n=AN_{11}=AN_{22}$. $A_n$ is the parameter in non-diagonal element of the mass matrix
and multiplies neutrino Yukawa coupling. So, it can give considerable effect to the cross section of direct detection.
As $A_n$ varies from $-1500$ GeV to $1500$ GeV, $\sigma^{SI}$ decreases from $1.2\times10^{-47}$ cm$^2$ to  $2.0\times10^{-48}$ cm$^2$.
 In the whole, it can satisfy the constraints from the direct detection experiments.

\section{discussion and conclusion}
BLMSSM is the extension of MSSM, and  we  add right-handed neutrinos, exotic Higgs singlets, exotic quarks and exotic leptons  to MSSM.
Through the seesaw mechanism, three light neutrinos  get tiny masses smaller than eV order.
 As we all know, the value of $\Omega_D h^2$ relates to  many parameters, such as $\tau_{ii}$ , $\rho_{ii}$ and $\mu_L$.
 Generally speaking, the parameters relating with the cross section of scalar neutrino annihilation can affect the results to some extent.
 We find reasonable
 parameter space to satisfy the relic density and direct detection experiments.
 In this condition, the lightest scalar neutrino mass is about 350 GeV as LSP in BLMSSM.

Now, theoretical physicists have come up with a lot of models for dark matter, and the types of dark matter candidates are also enriched.
In spite of this, the possibility space of reality is still much bigger than our imagine.
In our study, the lightest scalar neutrino could be a dark matter candidate in the reasonable range.
In fact, it is one of several possibilities.
Finally, we sincerely hope that people can uncover the real veil of dark matter as soon as possible through the full cooperation of physicists at home and abroad.

{\bf Acknowledgments}

We are very grateful to Jing-Jing Feng the Dr. of Beijing Normal University for solving some problems.
This work is supported by National Natural Science Foundation of China (NNSFC) 
(No. 11535002, No. 11605037, No. 11705045), Natural Science Foundation of Hebei Province (A 2020201002)
 and the youth top-notch talent support program of the Hebei Province.

\end{document}